\documentclass[trackchanges, 
twocolumn, 10pt]{aastex631}

\usepackage{graphicx}
\usepackage{ulem}
\usepackage{amsmath}
\usepackage[ruled,vlined]{algorithm2e}

\newcommand{\red}[1]{{\color{black} #1}}

\newcommand{\eg}{\emph{e.g.}}
\newcommand{\ie}{\emph{i.e.}}
\newcommand{\degree}{\ensuremath{^{\circ}}}
\newcommand{\SNR}{\ensuremath{\mathrm{\emph{SNR}}}}

\newcommand{\iou}{\ensuremath{\mathrm{\emph{IoU}}}}

\newcommand{\iop}{\ensuremath{\mathrm{\emph{IoP}}}}

\begin{document}

\title{Toward automated detection of light echoes in synoptic surveys: considerations on the application of the Deep Convolutional Neural Networks}






\correspondingauthor{Xiaolong Li}
\email{lixl@udel.edu}
\author[0000-0002-0514-5650]{Xiaolong Li}
\affiliation{Department of Physics and Astronomy, University of Delaware, Newark, DE 19716-2570, USA}

\author[0000-0002-8576-1487]{Federica B. Bianco}
\affiliation{Department of Physics and Astronomy, University of Delaware, Newark, DE 19716-2570, USA}
\affiliation{Biden School of Public Policy and Administration, University of Delaware, Newark, DE 19716-2570, USA}
\affiliation{Data Science Institute, University of Delaware, Newark, DE 19716-2570, USA}

\author[0000-0002-9276-3261]{Gregory Dobler}
\affiliation{Biden School of Public Policy and Administration, University of Delaware, Newark, DE 19716-2570, USA}
\affiliation{Department of Physics and Astronomy, University of Delaware, Newark, DE 19716-2570, USA}
\affiliation{Data Science Institute, University of Delaware, Newark, DE 19716-2570, USA}

\author{Roee Partoush}
\affiliation{Department of Physics and Astronomy,  Johns Hopkins University,
3400 North Charles Street, Baltimore, MD 21218, USA.}

\author[0000-0002-4410-5387]{Armin Rest}\affiliation{Space Telescope Science Institute, 3700 San Martin Dr., Baltimore, MD
21218, USA}\affiliation{Department of Physics and Astronomy,  Johns Hopkins University,
3400 North Charles Street, Baltimore, MD 21218, USA.}

\author[0000-0002-5947-2454]{Tatiana Acero-Cuellar }
\affiliation{Department of Physics and Astronomy, University of Delaware, Newark, DE 19716-2570, USA}
\affiliation{Observatorio Astronómico Nacional, Universidad Nacional de Colombia, Bogotá, Colombia}

\author{Riley Clarke}
\affiliation{Department of Physics and Astronomy, University of Delaware, Newark, DE 19716-2570, USA}

\author[0000-0001-7559-7890]{Willow Fox Fortino}
\affiliation{Department of Physics and Astronomy, University of Delaware, Newark, DE 19716-2570, USA}

\author{Somayeh Khakpash}
\affiliation{Department of Physics and Astronomy, University of Delaware, Newark, DE 19716-2570, USA}

\author{Ming Lian}
\affiliation{Department of Physics and Astronomy, University of Delaware, Newark, DE 19716-2570, USA}

\begin{abstract}
Light Echoes (LEs) are the reflections of astrophysical transients \red{off of} interstellar dust. They are fascinating astronomical phenomena that enable studies of the scattering dust as well as of the original transients. LEs, however, are rare and extremely difficult to detect as they appear as faint, diffuse, time-evolving features. The detection of LEs still largely relies on human inspection of images, a method unfeasible in the era of large synoptic surveys. The Vera C. Rubin Observatory Legacy Survey of Space and Time, LSST, will generate an unprecedented amount of astronomical imaging data at high spatial resolution, exquisite image quality, and over tens of thousands of square degrees of sky: an ideal survey for LEs. However, 
the Rubin data processing pipelines are optimized for the detection of point-sources and will entirely miss LEs. Over the past several years, Artificial Intelligence (AI) object detection frameworks have achieved and surpassed real-time, human-level performance. In this work, we prepare a dataset from the ATLAS telescope and test a popular AI object detection framework, \red{You Only Look Once, or YOLO}, \red{developed in the computer vision community,} to demonstrate the potential of AI in the detection of LEs \red{in astronomical images}. We find that an AI framework can reach human-level performance even with a size- and quality-limited dataset. We explore and highlight challenges, including  class imbalance and label incompleteness, and roadmap the work required to build an end-to-end pipeline for the automated detection and study of LEs in high-throughput astronomical surveys.  
\end{abstract}





\section{Introduction}
\label{sec:intro}

Light Echoes (LEs) are the reflections of light emitted by transients off \red{of} interstellar dust. While photons from a transient can reach us traveling along the line of sight, photons originally directed away from us can be reflected back to earth by favorably oriented dust sheets. The reflected image of the transient will inherit the complexity of the underlying dust structure, with morphological features at arcsecond or even arcminute spatial scales.

LEs provide invaluable information about the dust and the transient sources that originate them. They enable the study of dust and can reveal the dust structure around a transient in detail \citep{patat2005reflections}.  Through LEs, we have the unique opportunity to re-see ancient transients detected in direct light decades or even centuries ago, and study them with new technology and  instrumentation \citep{Rest11} so that LEs can be used, for example,  to classify historical supernovae \citep{rest2005light, rest2008scattered, rest2008spectral}. This was in fact how the Tycho and Cas A supernovae were classified \citep{rest2008spectral, krause2008cassiopeia, krause2008tycho}.  Finally, LEs can also provide a 3D view of an astronomical phenomenon, enabling the study of asymmetry of individual supernovae and supernova classes \citep{rest2011direct, finn2016comparison}. 

LEs are rare phenomena, as they require the serendipity of a bright transient and of the presence of a dust sheet oriented correctly to reflect the light towards the Earth. They are also extremely difficult to detect. The magnitude of a LE is $\sim10$ times fainter than its transient source \citep{patat2005reflections}. Thus, LEs bright enough for detection are extremely rare. \red{Furthermore}, the dynamic and complex dust environment makes their shape irregular and their identification more difficult than the identification of piont-source transients. \red{To date, no algorithm has proven successful in simultaneously detecting and localizing LEs in untargeted image surveys, and visual inspection of template subtracted images is the framework generally adopted for detection (see \autoref{sec:data})}.

The next generation of ground-based synoptic surveys, and in particular the Rubin Observatory Legacy Survey of Space and Time (LSST), will make visual-inspection of images unsuitable for practical purposes. Rubin LSST is scheduled to observe the sky continuously for 10 years starting in \red{2024}. The telescope will have an 8.4 m (6.5 m effective) primary mirror, a 9.6 deg$^2$ field of view, and a 3.2 Gigapixel camera which will produce 20Tb of information-dense data each night \citep{ivezic2019lsst}. LSST’s unique survey capability in the time domain and exquisite image quality will enable a vast and diverse range of scientific investigations. 
The LSST, with its unique combination of sensitivity, wide field of view, and dense observing cadence, has the capacity to revolutionize LE studies, pushing the field from novelty detections into statistically relevant regimes. But with nearly 1000 images each night for 10 years, it will become impossible to visually inspect the images to detect the LEs. An automated detection pipeline for LEs is needed to analyze the enormous outputs of this telescope. 

The LSST data processing pipeline under development is scoped to detect millions of alerts every night: anything that significantly changes (at a 5-$\sigma$ level) in the night sky from a reference image. However, the pipeline is designed for point-like sources, and it will entirely miss diffuse and extended transients, such as LEs. Our ultimate goal is to create an end-to-end pipeline for the detection and study of LEs in the LSST era. With this paper, we set the stage by exploring the potential of region-based Convolutional Neural Networks (CNNs) for LE discovery.

Automating the detection of LEs is a complex computer vision problem. Over the past several years, deep learning has proven successful in discovering intricate structures in high-dimensional data, \red{and} especially in image\red{s} \citep{lecun2015deep}. Deep learning-based object detection frameworks have achieved real-time, high-accuracy performance \citep{zhao2019object} that matches or surpasses that of humans. Deep Neural Networks (DNNs) have recently been used in the detection and study of astronomical sources, including the detection of galaxy clusters \citep{chan2019deep}, gravitational lenses \citep{davies2019using},  supernova remnants \citep{liu2019deep}, and more. \red{DNNs have also been applied to the detection of LEs in \cite{bhullar2021package}. This work produced a slide window detection model, ALED, which is based on a capsule CNN architecture \citep{sabour2017dynamic}. We compare this model in detail with our region-based model in the Results section of this paper (see \autoref{sec:results})}. 

In this paper, we present a proof-of-concept LE detection pipeline based on a \red{robust and proven} AI object detection framework, YOLOv3 \citep{redmon2018yolov3}, applied to the Asteroid Terrestrial-impact Last Alert System  \citep[ATLAS,][]{tonry2018atlas} telescope data. ATLAS provides a wide-field multi-epoch survey. Compared to the LSST, the survey for which we ultimately want to build a pipeline, the ATLAS survey is shallow (mag range 11–19.5) and coarse (resolution $~1.86''/$pixel); however, these data provide an excellent test-bed that allows us to demonstrate the potential of AI to detect LEs, and to explore the subtleties and nuances of architectures suitable to accomplish this task. 


\begin{figure*}
\centering
\includegraphics[scale=0.55]{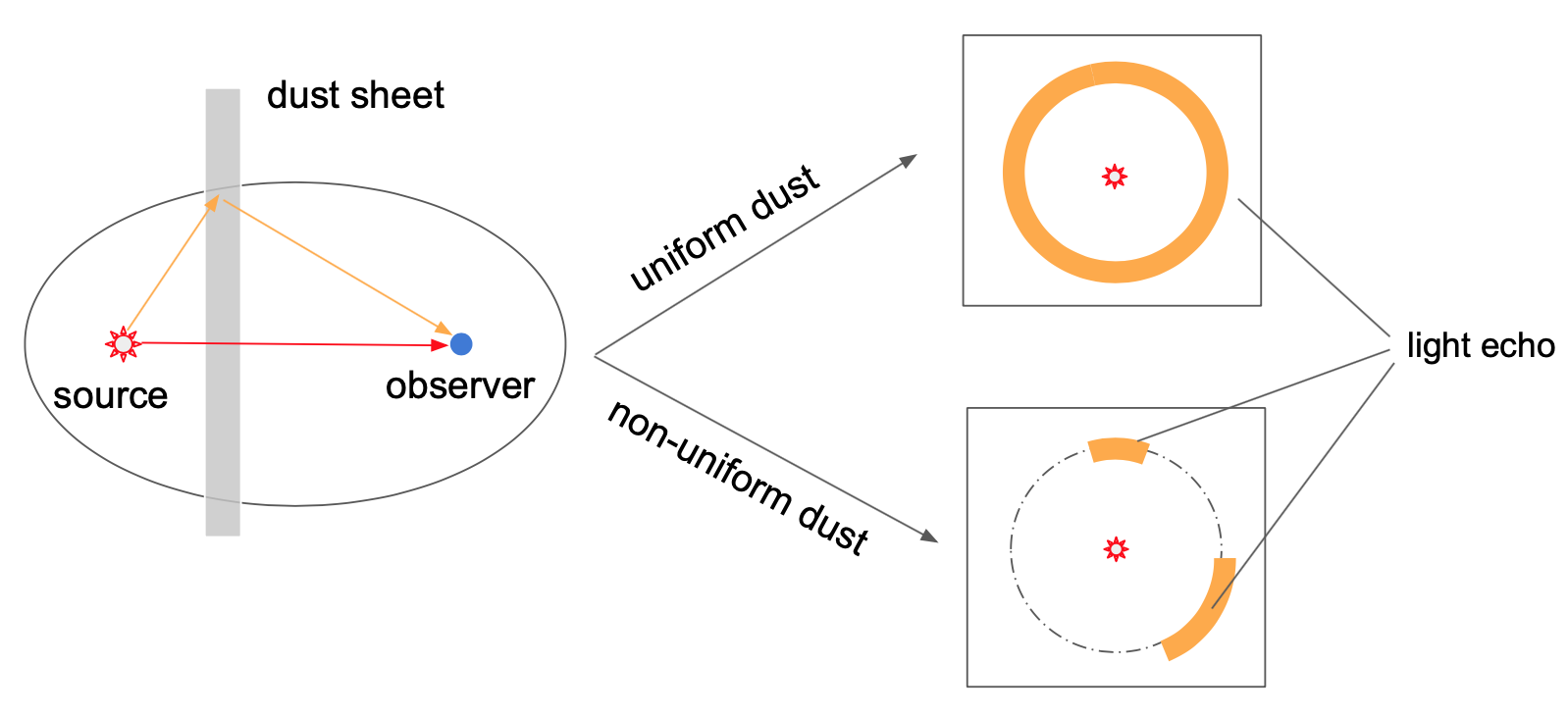}
\caption{A slice of the geometry of the LE  phenomenon (left) and its projection on the plane of the sky (right) 
in the case of \red{a} continuous dust sheet (top) and dust filaments (bottom). 
At each point in time, for every transient there is an ellipsoidal surface with foci \red{at} the earth and the transient. All points on the ellipsoid reflect light from the transient that arrive on earth at the same time and the ellipsoid itself expands over time. Due to the increase \red{in} path length, LEs reach the observer after the light from the original transient. On the plane of the sky, we see the cross-section of the reflection ellipsoid, yet not all LEs are circles: in most cases only a portion of the ellipsoid will have dust that reflects the light echo towards Earth. Most observed LEs are only arcs of a circle. 
}. 
\label{fig:LEgeo}
\end{figure*}

This paper is organized as follows.  We introduce the LE phenomenon and its history of discovery in \autoref{sec:history}. In \autoref{sec:data} we describe how our dataset is prepared. In \autoref{sec:method},  we present our detection framework, the definition of detection and the method of evaluation. The results of our model are discussed in  \autoref{sec:results}. Finally, in \autoref{sec:conclusion} we highlight challenges and the future work required to build an end-to-end automated pipeline for the detection and study of LEs.

\section{Historical note on Light Echoes}
\label{sec:history}
LEs were first discovered around Nova Persei 1901 \citep{ritchey1901changes,ritchey1902nebulosity}. Since then, LEs have been observed from a variety of sources, including variable stars: Galactic Nova Sagittarii \citep{swope1940notes}, Galactic Cepheid RS Puppis \citep{havlen1972nebulosity}, 
Nova Cygni \citep{bode1985dust}, 
OH 231.8+4.2 \citep{kastner1992variation, kastner1998direct},
V838 Monocerotis \citep{bond2003energetic}, 
the T Tauri star S CrA \citep{ortiz2010observation}, the Herbig Ae/Be star R CrA \citep{ortiz2010observation}; historical supernovae, of which we have centuries old records or that we only know from their remnants: SNR 0519-69.0 \citep{rest2005light}, SNR 0509-67.5 \citep{rest2005light}, SNR N103B \citep{rest2005light}, Cas A \citep{rest2007light, rest2008scattered, krause2008cassiopeia}, Tycho \citep{rest2007light, rest2008scattered, krause2008tycho}; modern supernovae, observed directly with modern astrophysical instrumentation as well, as in LEs: 1987A \citep{crotts1988discovery,suntzeff1988light}, 1980K \citep{sugerman2012thirty}, 1991T \citep{schmidt1994, sparks1999evolution}, 1993J \citep{sugerman2002multiple, liu2003scattered}, 1995E \citep{quinn2006light}, 1998bu\citep{ cappellaro2001detection},  1999ev \citep{maund2005hubble}, 2002hh \citep{welch2007extremely, otsuka2011late}, 2003gd \citep{sugerman2005discovery, van2006light, otsuka2011late}, 2004et  \citep{otsuka2011late}, 2006X \citep{wang2008detection, crotts2008nature}, 2006bc \citep{otsuka2011late, gallagher2012optical} 2006gy \citep{miller2010new}, 2007it \citep{andrews2011photometric}, 2007af \citep{drozdov2015detection}, 2008bk \citep{van2013echo}, 2012aw \citep{van2015legus}, 2014J \citep{crotts2015light, yang2017interstellar}, 2016adj \citep{sugerman2016discovery}; and the massive eruptive stellar system $\eta$ Carinae \citep{rest2012light, prieto2014light, smith2018exceptionally, smith2018light}.

The first systematic study of the phenomenon appeared in 1939 \citep{couderc1939aureoles}, and the theoretical framework was formalized further in  \cite{sugerman2003observability, tylenda2004light, patat2005reflections,patat2006reflections, Rest11}. As shown in \autoref{fig:LEgeo}, at each point in time, for any given transient there is an ellipsoidal surface with foci at the Earth and the transient.  If \red{there is} dust with the proper orientation on the ellipsoidal surface, photons that reach it are reflected towards the Earth. All points on the ellipsoid reflect light from the transient that arrives at the Earth at the same time \citep{patat2005reflections,patat2006reflections, Rest11}. Because of the increased path length, reflected light arrives at a delay compared to light traveling the direct path from the transient to the Earth. As time goes by, the ellipsoid expands and so does the travel path of LE light, giving us the opportunity to re-observe transients, even historical transients, that were directly observed centuries ago \citep{rest2005light, rest2007light, rest2008scattered, Rest11, rest2012light}.

In astronomical images, LEs appear as faint, morphologically diverse, diffused features. When they are reflected by dust directly surrounding the transient, circumstellar dust ejected from stellar eruptions for example, the LEs may appear as rings (see top plot of the right panel in \autoref{fig:LEgeo}). Most commonly, however, and when the source is a centuries-old transient, they take the shape of the dust filaments they reflect along a segment of the reflection ellipsoid (see bottom plot of the right panel in \autoref{fig:LEgeo}).

\section{Data} 
\label{sec:data}

In this work, we trained an AI model to detect LEs based on a dataset of visually inspected \emph{template subtracted images} from the ATLAS survey. 
\begin{figure*}[!t]
\centering
\includegraphics[scale=0.35]{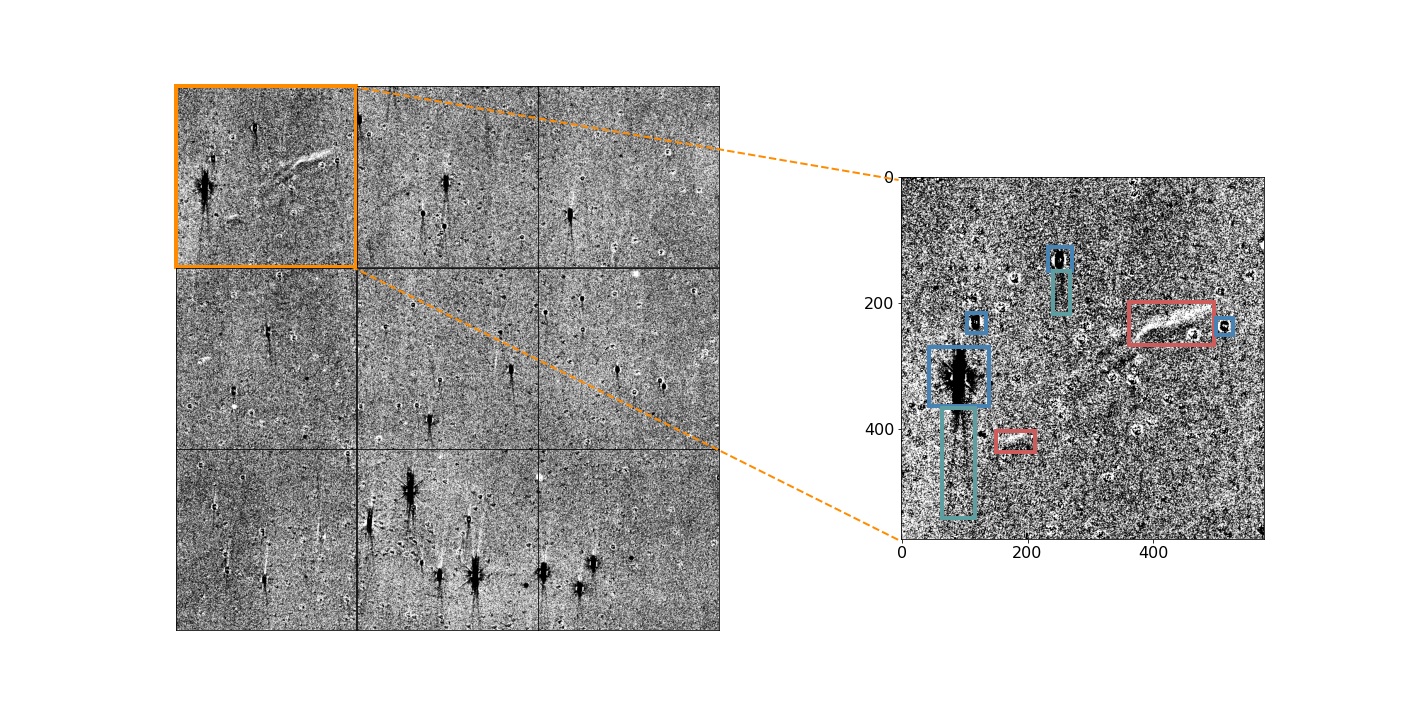}
\caption{An example of a template-subtracted (``difference'') image from the ATLAS survey (see \autoref{sec:data} for details). The native resolution of the difference image is $1800\times1800$ pixels, corresponding to about 1 square degree. The images are template-subtracted to aid detectability of the faint, time-evolving LE features.  
The full-frame image (left) contains a LE in the top left corner, as well as stars and artifacts throughout. 
Because of the rarity of LEs, in order to mitigate class imbalance between stars and LEs in our dataset, 
we split each ATLAS image into 9 tiles and select only the tiles that contains LEs. The top-left tile of this image that contains a LE is marked by an orange frame and \red{is} reproduced to the right with boxes corresponding to saturated stars, \red{star streaks, and LEs are shown in blue, teal, and red respectively}.
} \label{fig:LEsplit}
\end{figure*}

\begin{deluxetable}{ccc}
\tabletypesize{\footnotesize}
\tablewidth{0pt}
\tablecaption{ Technical specifications of the ATLAS system camera\label{tab:atlas}}
\tablehead{\colhead{Parameters} & \colhead{Value} }
\startdata 
Camera & Acam \\
Format (pixels) & 10560 $\times$ 10560 \\
Pixel Scale (arcsec) & 1.86   \\
FOV(degree) & 5.375 $\times$ 5.375\\
Number of filters & 7\\
\enddata
\tablecomments{Additional specifications are available in \url{https://atlas.fallingstar.com/specifications.php} }
\end{deluxetable}

\begin{deluxetable}{ccc}
\tabletypesize{\footnotesize}
\tablewidth{0pt}
\tablecaption{ LE data from ATLAS\label{tab:LEsearch}}
\tablehead{\colhead{Parameters} & \colhead{Value} }
\startdata 
RA & 0 ~ 360\\
DEC & -40 ~ 90\\
Epochs (MJD) & 58400, 58450, 58500, 58600, 58650 \\
Total number of images & 46800 x 5\\
\red{Filter} & $r$\\
Image size after DIA processing & 1800 $\times$1800 pixels (1 $\times$ 1 degree) \\
Images with LEs & 17 x 5 \\
\enddata

\end{deluxetable}

Template subtracted images can reveal transients that, due to their faint nature or the complex environment in which they arise (such as supernovae near the center of a Galaxy or a star forming region), are invisible to the naked eye in the original images. Template subtraction, or \red{Difference} Image Analysis (DIA), was developed as a technique for \red{detecting transients and monitoring microlensing events \citep{1989Natur.339..523N, 1992ApJ...399L..43C, 1995ASPC...77..297P}}.  Models based on the Optimal Image Subtraction (OIS) method \citep{alardlupton98} are \red{widely} deployed to survey-based searches for moving objects and changing astronomical phenomena. \red{Two images, of which one is considered a “template”, are aligned, PSF-matched, and subtracted from one-another.  The template itself can be a single, or more commonly a composite, image (generally from the same survey) with image properties that reach or approach the highest image quality the survey is capable of acquiring in terms of overall noise, spatial resolution, and depth.} After subtraction, the static background vanishes while any time-dependent phenomena, such as transients or moving objects, appear as a deviation from the 0-flux average. 

\begin{figure*}
\centering
\includegraphics[scale=0.35]{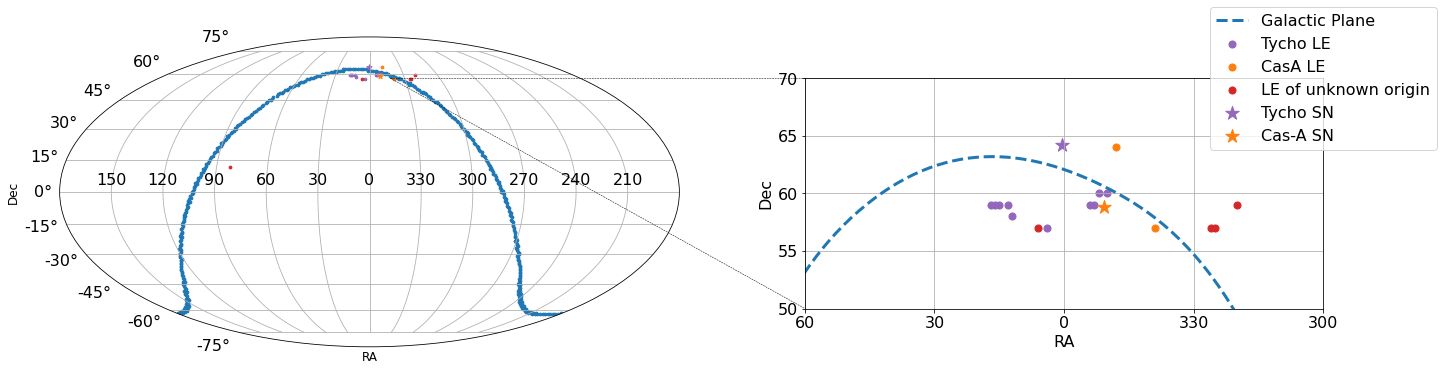}
\caption{The sky positions of 17 LEs contained in our final dataset. Ten of them are from the Tycho SN, and two \red{are} from Cas A. The rest have positions and orientations that are inconsistent with both Tycho and Cas A and are of as-of-yet unknown origin. 
LEs are preferentially found in high-dust content regions, such as near the Galactic plane, which is indicated by a thick blue line in this plot. The right panel shows a zoom-in view of the region containing the majority of the LEs recovered in our data. \autoref{tab:LEpos} contains the coordinate of each of the LEs in this figure.}
\label{fig:LEskymap}
\end{figure*}
\red{In practice, the DIA process is prone to the generation of artifacts that complicate the detection of transients in general, and LEs in particular.}

We assembled a dataset from template subtracted images from the ATLAS survey \citep{tonry2018atlas}.  The telescope has a $5.375\times 5.375$ degree field of view, \red{and the observational system is designed for the detection of asteroids}. \autoref{tab:atlas} shows the ATLAS system and survey characteristics, including the pixel resolution, sensitivity, filters used, and image size in pixels. \autoref{tab:LEsearch} shows the characteristics of the ATLAS data used in this paper.

Each image \red{produced by the ATLAS DIA pipeline} is originally  $1800\times1800$ pixels, covering a $\sim1\times 1$ degree field of the sky. \red{In the ATLAS difference images (\eg\ \autoref{fig:LEsplit}) we see several artifacts and imperfections produced by the DIA that complicate the detection of LEs: we notice residuals corresponding to saturated stars, with streaks typically associated with each saturated star. We also see residuals corresponding to stars below the saturation limit, due to sub-optimal image alignment or PSF matching.}

We visually inspected images within a sky region ranging from DEC $40\degree-90\degree$, RA $0\degree-360\degree$. This corresponds to about 234,000 inspected images as each field is observed at five different epochs \red{(see \autoref{tab:LEsearch})}. We select 17 \red{images that host LEs with sufficient signal-to-noise ratio (\SNR) to enable unambiguous detection by human visual inspection.\footnote{We note here that defining the \SNR\ of a LE is a complex task: the morphology is complex and the edges of the feature are not trivial to identify. We find that a functional definition of \SNR\ can exploit the characteristic ``dipole'' morphology. As the light travels across a sheet of dust, it illuminates different regions. In a difference image this results in clusters of pixel values positive on one side and negative on the other  (pointing in the direction of the original transient). Our functional definition of LE SNR then exploits the contrast in the dipole and the size of the region covered by the LE, measured through a simple segmentation procedure, as $SNR_\mathrm{LE}=IQR / log_{10}(N_\mathrm{pix})$. Ultimately, we find that this metric generally tracks the score our model assigns to our LEs.}} These LEs primarily originate from the Cas A and Tycho SNe, but we also found and included several LE-groups \red{in our training set } that are inconsistent with originating from either of those SNe and whose source is being investigated by our team. Note that hereafter we define a LE as a single contiguous dust filament lit by a transient's light, and a LE-group as an \red{ensemble} of individual LEs originating from a complex structure of dust with many filaments lit at the same time. Most LEs in our images are in LE-groups.  \autoref{fig:LEsplit} shows an example image from ATLAS with a LE-group composed of two LEs in the top-left corner. 


\autoref{fig:LEskymap} shows the sky-locations of the 17 LEs; their coordinates and transient sources are listed in \autoref{tab:LEpos}. \red{These constitute our dataset, the size of which is significantly smaller than those traditionally used to train region-based CNN object detection models in the computer vision literature (\eg, the original implementation of YOLO was trained on ImageNet, \citealt{russakovsky2015imagenet}, which consisted of more than 1,000,000 images with 1,000 different object classes)}.

\begin{deluxetable}{ccccc}
\tabletypesize{\footnotesize}
\tablewidth{0pt}
\tablecaption{Observed ATLAS LEs detected by visual inspection
\label{tab:LEpos}}
\tablehead{\colhead{RA} & \colhead{DEC} & \colhead{Gal l} &\colhead{Gal b} & \colhead{Source} }
\startdata 
4.0	&  57.0	 &  118.1 &  -5.5 & Tycho  \\
6.0	&  57.0	 &  119.2 &  -5.7 & unknown\\
12.0 & 58.0	 &  122.5 & -4.9  & Tycho  \\
13.0 & 59.0  &  123.0 &  -3.9 & Tycho \\
15.0 & 59.0	 &  124.0 &  -3.9 & Tycho  \\
16.0 & 59.0	 &  124.6 &  -3.8 & Tycho  \\
17.0 & 59.0	 &  125.1 &  -3.8 & Tycho  \\
82.0 & 12.0	 &  192.3 &  -12.4 & unknown \\
320.0 & 59.0 &  98.5 &   6.6 & unknown \\
325.0 & 57.0 &  99.1 &  3.3 & unknown \\
326.0 & 57.0 &  99.5 &  2.9 & unknown\\
339.0 & 57.0 &  105.3 & -1.2  & Cas A\\
348.0 & 64.0 &  112.3 &  3.2 & Cas A\\
350.0 & 60.0 &  111.7 &  -0.9 & Tycho\\
352.0 & 60.0 &  112.7 &  -1.2 & Tycho \\
353.0 & 59.0 &  112.8 &  -2.3 & Tycho \\
354.0 & 59.0 &  113.3 &  -2.5 & Tycho
\enddata
\tablecomments{The coordinates reported are for the difference image centroid.}
\end{deluxetable}

\begin{figure*}
\centering
\includegraphics[scale=0.4]{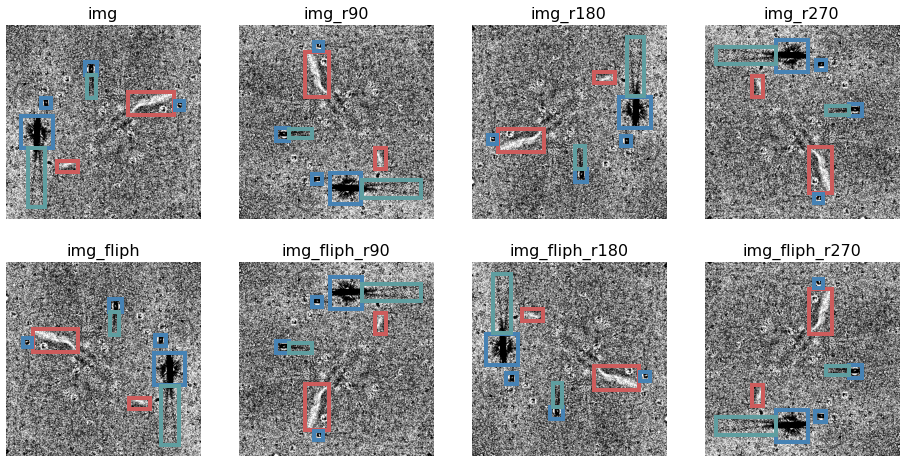}
\caption{Image augmentations applied to the ATLAS dataset: the images generated by subdividing the original ATLAS data are first flipped horizontally, and then rotated by 90, 180, \red{and} 270 degrees. 
\red{This} augmentation increases our dataset \red{size} by \red{a factor of 8}. 
}
\label{fig:LEaug}
\end{figure*}

In each image, we draw bounding boxes for \red{three} classes,  LEs,  stars \red{and  ``other'', which is a catch-all class for additional features and artifacts that may be present in an image, such as star streaks, satellites, \emph{etc}}. Notice that there is a certain amount of arbitrariness in the choice of the location and extent of a LE bounding box. Our bounding boxes strive to delimit each LE but end up, in many cases, including multiple LEs from the same LE-group. \red{We will return to this point when we describe our strategy for model evaluation in \autoref{ss:evaulation}}. 
We build our dataset in the COCO \citep{lin2014microsoft} format using \texttt{labelme}\footnote{\url{https://github.com/wkentaro/labelme}}. 

Because LEs are rare phenomena, much rarer than stars, 
there exists a large class imbalance between the number of objects in our \red{three} classes, and we note that imbalance in the training dataset can potentially introduce a bias in learning \citep{oksuz2020imbalance} and should therefore be addressed.  \red{The root of this potential bias lies in the fact that over-represented classes have more influence on the loss function, when the loss function is calculated as an average or median across all examples. Class imbalance in the training of machine learning models might thus lead to better performance on over-represented classes relative to under-represented ones.  However, we also note that in-class diversity and the SNR of individual objects are other important factors which may either reduce or enhance this effect. While we will implement a modification of the YOLO loss function that will help mitigate the risk of this bias (see \autoref{sec:method}),  we also address the class imbalance in data preparation.} We split the images into $576 \times 576$ pixel sub-images and only include in our dataset the image segments that host LEs.\footnote{YOLOv3 can accept input images at any size, \red{however} we choose 576 for \red{potential future work} since some CNNs model requires the input size to be \red{divisible} by 64.} 

We apply three data augmentation techniques to expand our dataset: we first flip each image horizontally, then rotate each image by $90\degree$, $180\degree$, and $270\degree$.  \autoref{fig:LEaug} shows an example of the results of image augmentation and \autoref{tab:LEaug} shows the number of images available in our dataset after each step of our augmentation process. The final dataset contains \red{224} images, \red{576} boxes labeled as LEs, and \red{1248} labeled as stars. 

\begin{figure*}
\centering

\includegraphics[scale=0.45]{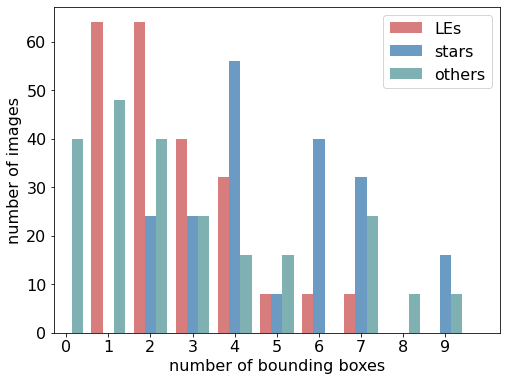}
\includegraphics[scale=0.45]{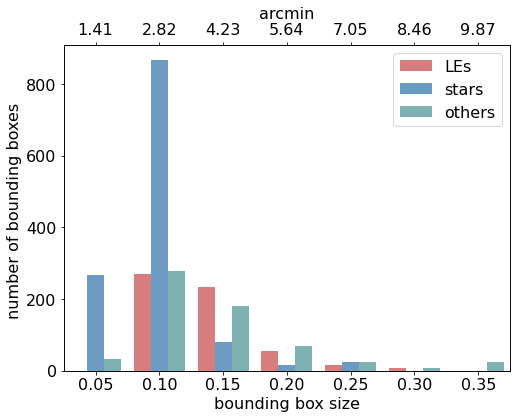}

\caption{The distribution of number of bounding boxes per image (left) and bounding box sizes across all images (right) for the two labeled classes: LEs, stars, and ``other'' (including satellite and star streaks and other artifacts). After augmentation (see \autoref{fig:LEaug}), the final dataset contains \red{224} images, \red{584} boxes labeled as LEs,  \red{1256} labeled as stars, and \red{648 as other}. On average, each image contains 3 LEs, 6 stars, \red{and} 3 ``other'' \red{objects}. The size of the bounding boxes is measured as the ratio of the diagonal of the box to the diagonal of the image. In these units, the mean size of LE and ``other'' boxes is $\sim0.1$ ($\sim 3~\mathrm{arcmin}$), while the stars are smaller, $\sim0.06$ ($\sim 1.8 ~\mathrm{arcmin}$).}
\label{fig:LEdist}
\end{figure*}

On average, each image contains three LEs, six stars, and four ``other'' objects. \autoref{fig:LEdist} shows the distribution of number of bounding boxes for both LEs and stars in each image, and the size distribution of the boxes. The size of the bounding box is measured as the ratio of the diagonal of the box and the diagonal of the image. The mean size of LE boxes (as well as ``other'' boxes) in these units is $\sim0.1$ ($\sim 3~\mathrm{arcmin}$), while the stars are smaller, $\sim0.06$ ($\sim 1.8~\mathrm{arcmin}$). 

Finally, the \red{pixel values in} images \red{that are fed into}  a neural network need to be appropriately scaled. Unlike normal 8-bit color RGB images, imaging data from the \red{ATLAS} telescope \red{is} stored as FITS files \citep{wells1979fits}. In our data, which is entirely comprised of template subtracted images, the FITS pixel values range from about -10000 to 5000. To feed the images as input to a traditional neural network architecture, however, a normalization is required. In order to constrain the pixel value range while minimizing the loss of information, we clip the raw FITS array to $[-10, 10]$, a range that contains about 95\% of the pixel values in our data. \red{We then standardize} the clipped images by subtracting the mean and dividing the result by the standard deviation \red{of the distribution of pixel values in the clipped FITS images}. 

Compared with other benchmark computer vision datasets, such as MS COCO \citep{lin2014microsoft} or Pascal VOC \citep{everingham2010pascal}, the dataset we have is far from ideal. First, as discussed earlier, the total number of images is small, even after augmentation. Second, the labels are incomplete. We labeled only two classes, while there are many kinds of objects contained in these images, such as satellites, artifacts, or star streaks, which may confuse the model.\red{We classify all these objects under the same ``other'' label}.  Third, the LE set is extremely small and all LEs are from a limited set of sources, primarily Tycho and Cas A, and may not span the \red{full} range of morphologies of the phenomenon.  
\red{However, } this dataset is a starter dataset for building a LE detection model, and used as a proof-of-concept here to explore the potential of AI applications to LE detection and the architectural modifications required to adapt existing region-based CNNs to this problem. 
\begin{deluxetable}{ccc}
\tabletypesize{\footnotesize}
\tablewidth{0pt}
\tablecaption{ Augmentation details\label{tab:LEaug}}
\tablehead{\colhead{Parameters} & \colhead{Value} & \colhead{Train/Test} } 
\startdata 
Initial number of images & 28 & 20/8\\
After flipping & 56 & 40/16\\
After rotating & 224 & 160/64\\
\enddata
\end{deluxetable}


\section{Methodology}
\label{sec:method}

\red{Beginning} in the late 1990s, the field of automated image recognition has been dominated by a family of AI models called Convolutional Neural Networks  \citep[CNNs, ][]{lecun1989backpropagation}. Today, CNNs \red{-- alongside ``attention''-based ``transformer'' models \citep[e.g.,][]{vaswani2017attention} --} remain \red{one of} the primary \red{families} of models for image recognition tasks. Object detection frameworks based on CNNs have achieved and surpassed human performance. There are two main CNN frameworks for \red{task of detecting \citep{zhao2019object} and localizing objects in images (i.e., region-based CNNs): region proposal-based two-stage methods that first propose regions with one CNN and an object in those regions with another \citep{girshick2014rich,girshick2015fast,ren2015faster} and single-stage methods \citep[e.g.,][]{redmon2016you} 
that identify and classify regions hrough a single deep CNN model}. 
In this paper we focus on single-stage CNN methods, using \red{the third publicly available version of the \emph{``You Only Look Once''} (YOLOv3) model \citep{redmon2018yolov3}}.

A few choices have to be made and challenges have to be faced for \red{the} detection of LEs which are unlike those traditionally encountered when detecting common computer vision targets in the field of object identification. These challenges are shared by other fields of research (\eg\ other applications of AI to astrophysical tasks, or in the medical field) but in combination they may be unique or at least rare.
\begin{itemize}

\item
LEs are not finite objects: they exist at all levels of SNR and their edges blend into the background as the dust that reflects them blends into empty space and the light they reflect fades. Thus we cannot aspire to be complete in our human labeling because where LEs blend into the noise human detection fails. Since our training set is not complete, we have to teach the CNN both what is and what is not a LE: a multi-label model. In this initial work we only \red{train} the model to \red{identify} LEs, stars, and a limited number of artifacts.

\item
Of the many features in an image, LEs are an extremely rare occurrence. Stars and galaxies, transients and variable phenomena, moving objects, and several artifacts overwhelm the image searches. Detection of rare features is a problem often encountered in medical applications of computer vision \citep[][and references therein]{sanchez2020deep}. 
We explore the detection of LEs in a subset of our data where each image contains at least one LE (as described in \autoref{sec:data}) because we want to focus on the architectural choices that enable the detection of LE features. 

\item
Putting bounding boxes around LEs requires us to make choices that are, to some extent, arbitrary: each LE inherits a complex structure from the underlying dust and most LE filaments are connected into complex structures (LE-group, \autoref{sec:data}). While we strive for consistency in setting the bounding boxes around isolated filaments, we cannot expect that the model will generate bounding boxes that exactly overlap each one of the training bounding boxed. We address this issue by modifying the region-based CNN loss (see \autoref{ss:yolov3}) exploring  various approaches to mitigating this problem, including the application of focal loss \citep{lin2017focal, vaswani2017attention}, and with a specific  definition of True and \red{F}alse positives (see \autoref{ss:evaulation}) \red{that differ in the training and evaluation phases}.
\end{itemize}

\subsection{YOLOv3}
\label{ss:yolov3}

\begin{figure*}
\centering
\includegraphics[scale=0.6]{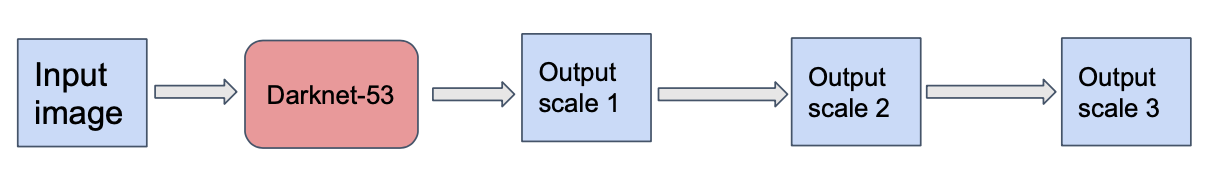}
\caption{Architecture of the YOLOv3. An input image is mapped to a tensor of shape $S\times S \times N_b \times (5 + N_c)$ through a deep neural network model as described in \autoref{ss:yolov3}. Each cell of the $S\times S$ grid is responsible for both class prediction and bounding boxes regression of objects whose centers fall inside the cell. The YOLOv3 model makes predictions at three size scales (1/8, 1/16, 1/32 of the image size). }
\label{fig:yolov3}
\end{figure*}

YOLO \citep[You Only Look Once][]{redmon2018yolov3} is a popular \red{single}-stage regression/classification region-based framework \red{where ``regression'' refers to the identification of regions of interest through regression of bounding box coordinates and ``classification'' refers to the classification of the object inside of those bounding box regions}. Feature extraction and object localization are integrated into a single deep neural network model. An input image is split into an $S \times S$ grid of cells. Each cell is responsible for both class prediction and bounding boxes regression of objects whose centers fall inside the cell. In other words, through a \red{DNN} model, an input image is mapped to a tensor with the shape $S \times S \times N_b \times (5 + N_c)$, where $N_b$ is the number of predefined anchor boxes assigned to each cell, \red{each} containing the  probabilities $p_c$ for each of the $N_c$ classes. The number 5 comes from having one parameter $C$ for the confidence that an object is present in the bounding box and four parameters to define the position and size of the bounding box $b_x, b_y, b_h, b_w$. 

In this paper we adopt the architecture of YOLOv3 \citep{redmon2018yolov3}.  \red{We note that our choice of YOLOv3 is motivated by a desire to use a model that is both ``foundational'',  in terms of region-based CNN object detection methodology (\citealt{girshick2014rich,girshick2015fast,ren2015faster,redmon2016you}, see also \citealt{sultana2020review} for a review), stable, and with comparable performance to models under active development.  That is, one of the goals of this paper is to demonstrate that region-based CNN models can be used for the purpose of detecting LEs in upcoming surveys, and so future work can be aimed at assessing the relative advantages of one particular implementation over another.  The YOLOv3 architecture is} shown in \autoref{fig:yolov3}. YOLOv3 uses Darknet-53, which consists of 53 deep convolutional layers, as the backbone network and makes predictions at three scales (1/8, 1/16, and 1/32 the size of the input image).\footnote{As pointed out in \cite{redmon2018yolov3}, Darknet-53 is a powerful model which has comparable accuracy to ResNet-101 \citep{he2016deep} used in Faster RCNN \citep{ren2015faster}, but the speed is 1.5$\times$ faster. Our preliminary work comparing results based on YOLO with results obtained from Faster RCNN does not suggest that the one-stage YOLO architecture is less suitable for the detection of LEs than a more traditional two-stage RCNN model when comparing model accuracies for the two.}  
\red{We pass our $576 \times 576$ ATLAS images (each of which contains at least one light echo as described above) through Darknet-53 which, because its architecture includes multiple convolutional layers with strides equal to 2, results in a downsampling to three scales appropriate for the detection of LEs in our sample, $72\times 72$, $36\times36$, and $18\times 18$ pixels.  These feature maps are passed to the classifier and bounding box regression network which, for each of these scales,  assigns three anchor boxes (regions) of different axis ratio to each grid cell, resulting in $(72\times 72 + 36\times36 + 18\times 18) \times 3 = 20412$ bounding boxes. The regression network assigns a score for each class to each set of the bounding boxes' coordinates,  
generating predictions for those 20412 boxes.} 

The predicted bounding boxes are first filtered by thresholding on the  probabilistic classification score $p_s$, which is the confidence $C$ times the probability $p_c$ that the object is of a certain class, $p_s = C p_c$, in order to eliminate boxes with a low score. Non-maximum suppression \citep[NMS][]{neubeck2006efficient} is then applied to keep only one box when several boxes overlap with each other. 

YOLOv3 minimizes a loss function which consists of three parts: confidence loss $L_{conf}$, classification loss $L_{cls}$, and bounding box regression loss $L_{box}$,
\begin{equation}
    L_{total} =  L_{conf} + L_{cls} + L_{box}.
\label{yolo_loss}
\end{equation}
\red{The confidence loss $L_{conf}$ represents the loss from the confidence probability $C$ (1 if there is an object in the bounding box and 0 for background), while} $L_{cls}$ measures the ability make a correct classification. We use a simple cross entropy loss,
\begin{equation}
    L_{cls} = \sum_{i=0}^{S^2} \sum_{j=0}^{B} I_{ij}^{obj} f_{CE}( p_{ij},  \hat{p}_{ij})
\end{equation}
where $S^2$ denotes the three scales of \red{the} anchor boxes, and $B$ is the total number of anchor boxes assigned to each grid cell. $I_{ij}^{obj}$ denotes whether there is a object in the cell (it is 1 when there is an object and 0 otherwise), \red{while} $f_{CE}$ stands for the cross entropy function given by:
\begin{equation}
    f_{CE}(x, y) = -y\log(x) - (1-y)\log(1-x).
\label{eq:cross_entropy}
\end{equation}

The number of background anchor boxes is typically much larger than the number of foreground anchor boxes. To address this problem, \citet{lin2017focal} proposed to use the ``focal'' loss, which adds a factor to to the cross entropy loss. We adopt a similar approach for the detection of LEs, since as shown in \autoref{sec:data}, on average there are only 3 LEs in each image. We redesign $L_{conf}$, 
\begin{equation}
\begin{split}
 L_{conf} =  \sum_{i=0}^{S^2} \sum_{j=0}^{B} I_{ij}^{obj} (1-C_{ij})^\gamma f_{CE}( C_{ij}, \hat{C}_{ij} ) \\
    + \sum_{i=0}^{S^2} \sum_{j=0}^{B} I_{ij}^{no-obj} (1-C_{ij})^\gamma f_{CE}( C_{ij}, \hat{C}_{ij} )
\end{split}
\end{equation}
where $I_{ij}^{obj}$ and $I_{ij}^{no-obj}$ are mask factor\red{s that are the} same as in $L_{cls}$, except $I_{ij}^{no-obj}$ is 1 when there are no objects in the cell. $f_{CE}$ is the cross entropy function defined in \autoref{eq:cross_entropy} and $(1-C_{ij})^\gamma$ is the focal loss term.  We set the value $\gamma=2$ as suggested in \cite{lin2017focal}.

$L_{box}$ quantifies the coordinate difference between predicted and labeled bounding boxes. The most commonly used method for comparing the similarity between two bounding boxes is by computing the Intersection over Union (\iou), the ratio between the area of the intersection and the area of the union of two boxes. The more overlap, the greater the \iou, the more closely aligned are the two boxes. Then the loss based on \iou\ is just $L_{\iou}= 1-IoU$. In YOLOv3, the \iou\ loss is weighted by a factor related to the area of the boxes to improve the performance for small boxes. Overall,
\begin{equation}
    L_{box} = \sum_{i=0}^{S^2} \sum_{j=0}^{B}I_{ij}^{obj} (1-IoU) (2 -  \hat{w}_{ij} \hat{h}_{ij})
\end{equation}
where $\hat{w}_{ij}$ and $\hat{h}_{ij}$ are the width and height of the labeled box.


However, \red{as we have described above}, it is particularly tricky to label the LEs in an unambiguous way. Inside of a label box there are \red{often} multiple LEs and how to group the LEs is somewhat arbitrary. In the face of this ambiguity, the \iou\ loss is not able to properly reflect the properties of LEs and a traditional \iou\ may penalize good detections that split a LE-group differently than a visual inspector, but in just as valid a way. To enable the model to learn features inside the box and avoid penalizing fair choices that happen to be different than the ones implemented by the human labeler, we replace the \iou\ with the \red{\iop, intersection over prediction (where prediction is the area of the annotation label), in the evaluation phase. Simultaneously we generate larger annotation boxes for evaluation. In the training phase, we prepare annotations designed to separate each LE in a group. In evaluation, however, we use annotations designed to include all LEs in a LE-group, so that a different segmentation of the LE-group would not be penalized by our model.}

\begin{deluxetable}{ccc}
\tabletypesize{\footnotesize}
\tablewidth{0pt}
\tablecaption{ Hyperparameters of YOLOv3\label{tab:yolo}}
\tablehead{\colhead{Parameters} & \colhead{Value} }
\startdata 
Backbone & Darknet-53 \\
Feature Strides &  [8, 16, 32]\\
Anchor per scale &  3\\
Initial learning rate & $10^{-4}$ \\
Final learning rate & $10^{-6}$
\enddata
\end{deluxetable}


\subsection{Detection definition and Evaluation} \label{ss:evaulation}

While the loss minimization \red{determines} the parameter \red{values} of a model, some parameters have to be set by the user to select the optimal model \red{for object detection}. In the case of our CNN, in addition to the hyperparameter choices indicated in \autoref{tab:yolo}, this means choosing the thresholds for detection and classification above which a detection is considered valid. There are two thresholds that \red{need to} be chosen: 
\begin{itemize}
\item the probabilistic classification score $p_s$ \red{threshold},
\item the \iou\ threshold.
\end{itemize}
To evaluate the best threshold values we need to define what True/False Positives are based on the overlap of a label and a bounding box and the probabilistic classification of the object inside the box.
We consider a detected bounding box as a true positive (TP) if 
\begin{itemize}
    \item the class predicted is the same as labeled (star or LE) \emph{and} with score $p_s$ larger than a threshold;
    \item the overlap between predicted and ground truth bounding boxes measured as their intersection over the union is larger than a threshold \iou. 
\end{itemize}
    
Conversely, we count it as a false positive (FP), if
 \begin{itemize}
 \item class id is the same as labeled with $p_s$ larger than a threshold, but \iou\ is less than the threshold.
 \end{itemize}

\begin{figure*}
\centering
\includegraphics[scale=0.32]{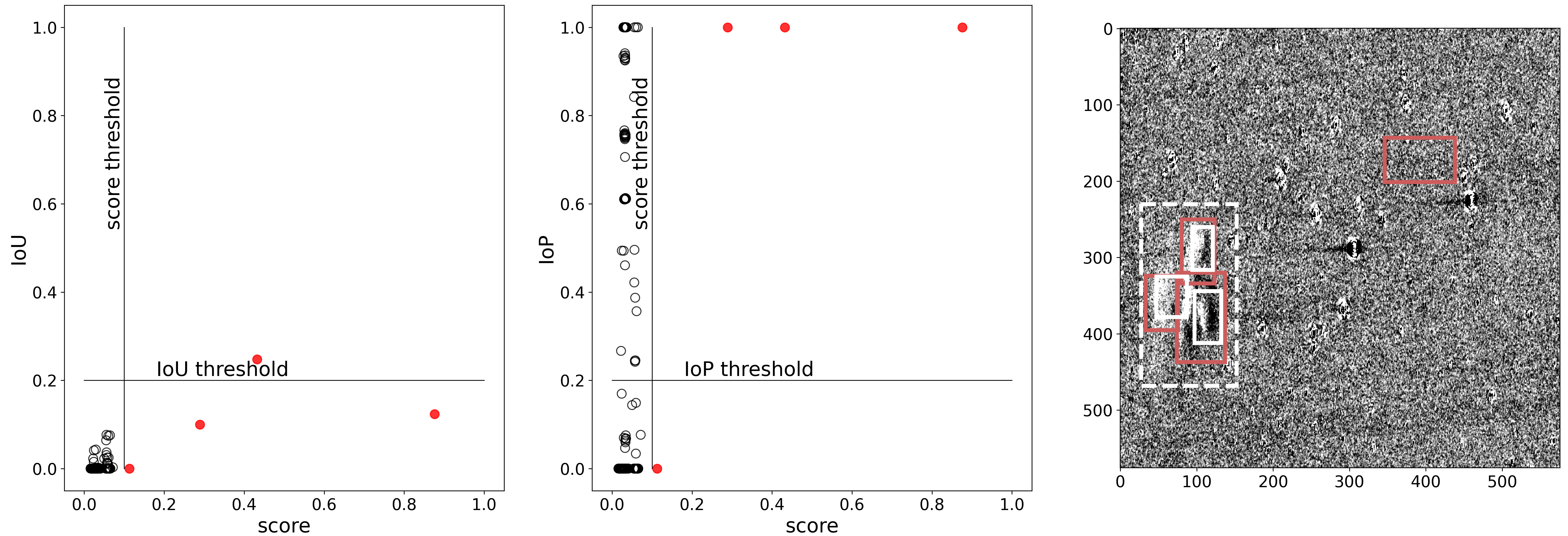}
\caption{\red{Predictions from YOLOv3 for an example image from our test set. The left panel shows a scatter plot of classification score vs. \iou\ ---intersection over union--- the traditional metric used in YOLO for training and evaluation, for all predicted bounding boxes. Four boxes pass the score threshold, set to 0.1; those are indicated by red circles in the left and center panels of this figure. However, the \iou\ value is $\iou<0.2$, our chosen threshold, for three out of the four boxes. Based on the traditional \iou\ definition, three of these detections would be considered FPs.  Due to the complexity of the LE morphology, the label boxes are somewhat arbitrary in their size and position and LEs could be split in different ways. In the right-most figure we mark the labels that we use in training phase (solid white boxes, typically smaller) and the labels that we use for model evaluation, that are designed to contain a light echo group in its entirety (white dashed box). While three of the four detection boxes do not pass the \iou\ threshold, two of them are in fact contained entirely inside of an evaluation label box. We calculate the \iop\ as the intersection of prediction and evaluation label over the area of the predicted box and replace the \iou\ with the \iop\ for model evaluation. In this example, the model successfully detected the LE group with three detections, all above the \iop\ threshold and this image contains only one FP, for which $\iop=0$.}}
\label{fig:yolo_pred}
\end{figure*}

Based on the properties of LEs discussed above, \red{the standard \iou\ is not a good evaluation metric for LEs. The LEs are extended features without clear boundaries, and valid prediction boxes that fall within an annotation box would have low \iou.  Taking this into consideration, when searching for LEs, we replace the \iou\  with the \iop for LEs, while for the stars and ``other'' category we continue to use the standard \iou\ as the location and boundaries of the bounding box are generally not ambiguous. This reduces the number of False Negatives (FNs).  Finally, predicted boxes that overlap with the same label box (see \autoref{fig:yolo_pred}) need to be consolidated so as to not to double count TPs.}

Once we define the TP and FP, we are able \red{to determine} the receiver operating characteristic (ROC) curve, \red{which indicates the number of TPs vs FPs at different thresholds. The two-dimensional area underneath the curve (AUC) measures the overall performance of the model: the larger the area, the better the model.  In addition, we can also then calculate} the precision-recall (PR) curve \red{for various } \iou\ (or \iop\ for LEs) \red{and} classification score \red{thresholds}. 

Precision and recall are defined as 
\begin{equation}
    P = \frac{TP}{TP+FP} = \frac{TP}{N_{det}}
\end{equation}
\begin{equation}
R = \frac{TP}{TP+FN} = \frac{TP}{N_{ann}}
\end{equation}
where $N_{det}$ is the total number of detected boxes, and $N_{ann}$ is the total number of labeled boxes.  Precision measures the fraction of positive detections that are correctly classified, while recall measures the completeness fraction. Precision and recall are often in tension: improving precision by increasing the classification and/or detection threshold, typically reduces recall, and vice versa. Therefore, the F1 score, defined as the harmonic mean of precision and recall $F_1 = 2 \frac{PR}{P+R}$, is often used as a measure of balance between precision and recall. The evaluation steps \red{for our models} are summarized in \autoref{algo:pr}.


\begin{algorithm}
\SetAlgoLined
 $N_{ann} \gets$ number of labeled objects \;
 $N_{det} \gets$ number of predicted bounding boxes\;
  set classification score threshold \;
  set \iop\ detection threshold\;
  $N_{TP} \gets 0$\;
  $N_{FP} \gets 0$\;
    
  \For {each predicted bounding box}
  {\If {classification score $>$  threshold} 
        {\If {\iop\ $>$ \iop\ threshold} 
            {$N_{TP} \gets N_{TP}+1$ }
        \Else
            { $N_{FP} \gets N_{FP}+1$ }
       }
}
   $P = N_{TP} / N_{det} $, 
   
   $R = N_{TP} / N_{ann} $,
   
   $F1 = 2 \frac{PR}{P+R}$
\caption{LE detection model evaluation}
\label{algo:pr}
\end{algorithm}


\subsection{Training} \label{ss:training}

We trained our model on the \texttt{Google Colaboratory} platform. \red{We adopt a three-fold cross validation scheme in which 28 images containing LEs were randomly split into 10, 10 and 8 subsets. We kept one subset as a test set, and trained the model on the other two subsets. We shuffled the images once and repeated this process, thus obtaining six cross-validation sets. The results we present in \autoref{sec:results} represent the average of the results across these six cross-validation sets.}  We used an Adam optimizer \citep{kingma2014adam} with an adaptive learning rate, setting the initial and final values to $10^{-4}$ and $10^{-6}$ respectively. We applied a learning rate schedule to improve the training progress as described in \cite {he2019bag}. The learning rate is decreased from initial value following a cosine function.


\section{Results}
\label{sec:results}
\begin{figure}
\centering
\includegraphics[scale=0.38]{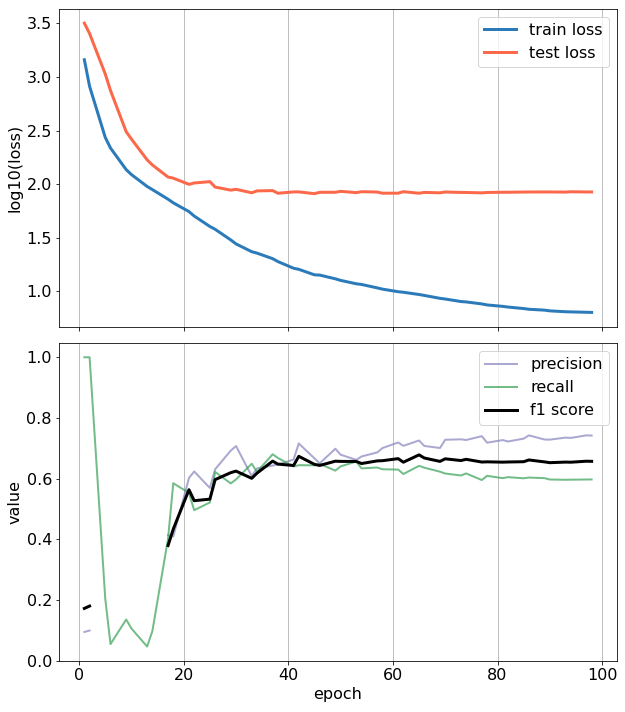}
\caption{Training of YOLOv3, discussed in \autoref{ss:training}. \red{All curves in this figure represent averages across the 6 cross-validation datasets}. The top panel shows the total loss for the training and test set, as labeled. The total loss is defined for YOLOv3 as the sum of classification and regression losses (see \autoref{ss:yolov3}). The test loss shows little improvement after about 50 epochs. The bottom plot shows the precision and recall for LEs as a function of training epoch, as well as the \emph{F1}-score, as labeled. Precision and recall are calculated with a classification score threshold of 0.2 and \red{\iop\ threshold of 0.2}. 
The \emph{F1}-score is the harmonic mean of precision and recall, demonstrating how a balance between the two is achieved near 50 epochs.}
\label{fig:yolo_loss}
\end{figure}

\red{\autoref{fig:yolo_loss} shows our performance over multiple training epochs averaged over the six cross-validation sets.}. The top panel in \red{the figure} shows the total loss  (\autoref{yolo_loss}) for the training and test sets. The training loss continues to decrease throughout all epochs, while \red{the} test loss reaches \red{a} minimum around 50 epochs and then rises slowly due to overfitting. The precision and recall for LEs on the test set are shown in the bottom panel. Precision and recall are calculated with a classification score threshold of 0.2 and \iop\ threshold of 0.1. 
After \red{$\sim$}50 epochs, the $F1$ score stabilizes, and we choose the model state at 50 epochs as the final model for further evaluation.

\autoref{fig:yolo_pred} shows an example of \red{the} YOLO predictions from our test set. Setting the score threshold to 0.1, there are four boxes above threshold. All but one of the four boxes have \iou\ values that are less than 0.1 and would be considered as FPs. But they are actually good detections: the model successfully detected several LEs belonging to the same LE-group within the labeled bounding box. \red{Thus, the \iop\ replaces the \iou\ for LEs in the evaluation phase. Note that we use the \iop\  only for LEs.  For the stars and ``other'' classes we use the \iou\ since the prediction box annotations are much less unambiguous, and the predictions} are never in fact contained in the label boxes, so replacing the \iou\ is unnecessary.

\begin{figure*}
\centering
\gridline{
\fig{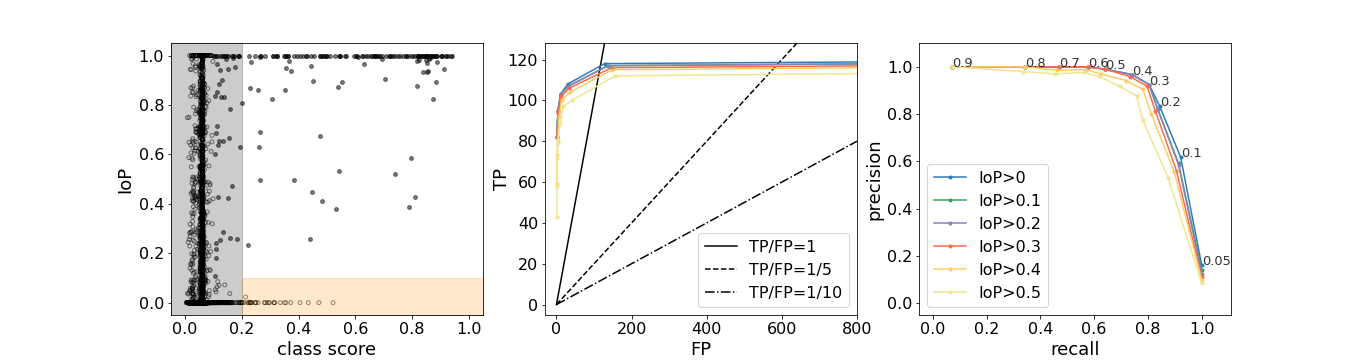}{0.95\textwidth}{(a)} }
\gridline{
\fig{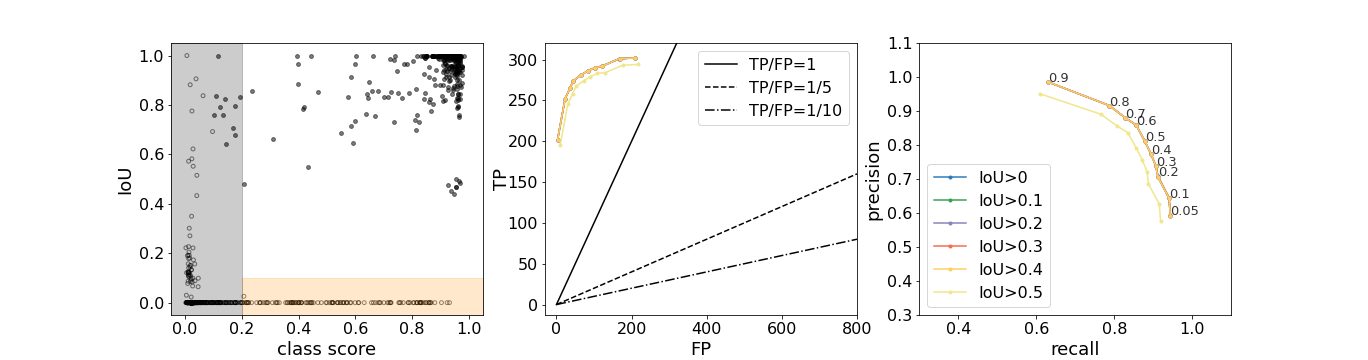}{0.95\textwidth}{(b)}
}
\caption{Test results from YOLOv3 for LEs (a) and stars (b) for a chosen cross-validation set. In both (a) and (b), the left panel shows a scatter plot of classification score vs. \iop\ (top, for LEs) or \iou\ (bottom, for stars) for all predicted bounding boxes. The middle panels show ROC curves and the right panels show PR curves. The grey regions in the left plots delimit the region of rejected predictions (below a score threshold of 0.1) and the orange regions represent the location where FPs are found (\iou\ or \iop\ smaller than 0.2). 
The ROC and PR are plotted for different thresholds of score (points along each curve) and different \iou\ or \iop\ (as labeled). In the ROC curve plots, the three black lines indicate a ratio of true positives to FPs of 1/1, 1/5 and 1/10. In the PR curve, gray lines join values obtained for the same score for different \iop\ thresholds.  
\red{If we choose an $\iop>0.1$ and a classification score threshold of $0.4$, larger than 70\%  recall is achieved at a precision of about 95\%. This means that we have to tolerate a 5\% percent FPs to detect 70\% of LEs.} Because of the limited size of our test set, we leave studies of the influence of size, noise level and brightness of LEs on the precision/recall to future work. 
}
\label{fig:yolo_roc}
\end{figure*}

The ROC and PR curves show the model performance for a given box-overlap threshold as the threshold on the classification probability is modified. The ROC and PR curves for \red{\iop\ (\iou) thresholds for LEs (stars) from 0 to 0.5 are shown in \autoref{fig:yolo_roc}.  The top panel in \autoref{fig:yolo_roc} shows that, for this particular cross-validation set, if we choose an $\iop>0.1$, $R>70\%$ recall is achieved at a precision of about $P\sim95\%$ (score=0.4).  This means that we have to tolerate a 5\% FP rate to detect at least 70\% of the LEs.

We note that both probability classification threshold and box-overlap threshold have to be set to very low values to achieve the desired performance (0.2 and 0.1 respectively for our best preforming model). This implies that the classification score shall not be interpreted as a ``probability'' without performing a re-calibration; the classification score for stars and for LEs cannot be compared and thresholds should be set separately for each class. }

\begin{figure*}
\centering
\gridline{
\fig{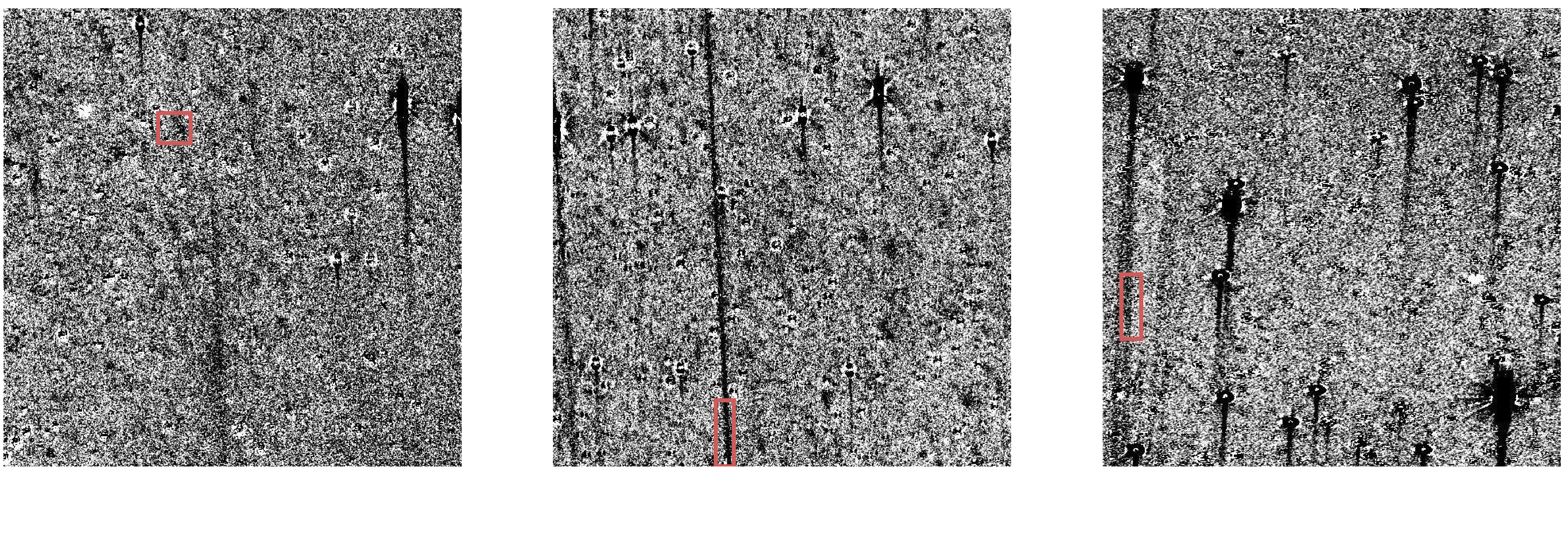}{0.60\textwidth}{(a)} 
\fig{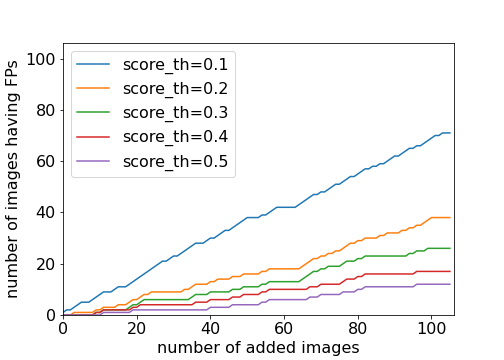}{0.42\textwidth}{(b)}
}
\caption{{(a) Three types of common FPs from our implementation of YOLOv3 \red{for LE searches in ATLAS images} for the same cross-validation set used in \autoref{fig:yolo_roc}. FPs are primarily associated with two kinds of features in the images.  ``Dipole'' structures, \ie\ features in the difference image that show one bright and one dark side corresponding to moving objects, including LEs, but also, for example, asteroids generate FPs. This kind of FP is shown in the left most image in panel (a). The second kind of FPs are generated by saturated star streaks --- middle and right images in panel (a). \red{The class ``other'' is designed to train the model to recognize non LE artifacts, including these features. While this class suppresses the number of FPs compared to models trained on two classes only (LE and star), some of these artifacts are still incorrectly classified as LEs.  Panel (b) shows that the rate of growth of FPs as a function of the number of images without LEs for which the model is asked to make prediction grows sublinearly for each of five score threshols between 0.1 and 0.5.  This figure is discussed in \autoref{sec:results}}.
}}
\label{fig:false_positive}
\end{figure*}

With the large amount of astrophysical images to be produced with our target survey, LSST, we must control FPs in our model or else we will be overwhelmed with an unmanageable number of images to visually inspect.  Or worse, we may deploy and waste follow up resources if we trust the automated detections without vetting them.  As indicated in \autoref{sec:data}, we prepared a training set that contains LEs in every image, but LEs are rare events and the vast majority of images from any surveys contain none. To test the robustness of the model, we added 100 images to the test set that contain stars, potentially artifacts, but no LEs. Following the same evaluation methods described in \autoref{ss:evaulation}, we found that precision drops with the score threshold. The rate of FPs grows sublinearly with the number of added images. The slope at score 0.5 is about 0.1, while at score 0.2, it is 0.3. We also found that these FPs largely appear near the edge of stars or where dipole structures, structures associated with poor image subtraction due to mismatched PSF between template and search image, or poor local alignment of the image, are present, see \autoref{fig:false_positive}.




\red{Both our model and ALED \citep{bhullar2021package}, the only other automated LE detection method we are aware of, can achieves a high ($P\sim90\%$) precision and high recall  ($R\sim90\%$) although, since the models were applied to different datasets, it is not straightforward to compare their performance at this time. We note that the final results reported for our model are on the median across the 6 cross-validation sets, while it appears that ALED did not perform cross validation. On a single set, our precision and recall can be as high as 0.9 simultaneously (see \autoref{fig:yolo_roc}), but perhaps the most fair comparison is that with the model reported in table 4 of \cite{bhullar2021package} as ``ALED-m'': a precision $P=1$, and a recall $R=0.25$. When taking the median across all cross-validation sets, at $P_\mathrm{median}=1$ we measure $R_\mathrm{median}$ = 0.3 for score and \iop\ thresholds of 0.7 and 0.2 respectively.  While both datasets used for training the models are small, our dataset contains only 28 576$\times$576 images with LEs (before augmentation) compared to 175 200$\times$200 images in \cite{bhullar2021package}. There are significant differences in the methodological approach. First, the ALED model identifies the presence of a light echo in an image (via binary classification) but does not locate the LE directly. The YOLO framework performs the detection and localization simultaneously, and returns class type as well as the exact coordinates of positions which is important if we envision further automation of the LE study pipeline, such as automatic selection of LEs for follow-up and automatic follow-up on robotic systems, an increasingly common practice in astrophysics (\eg, \citealt{2018SPIE10707E..11S}). We note, however, that the location of the LEs in the images can be estimated in a follow up step through the analysis of the ALED feature maps. In addition, the training of ALED requires a balanced dataset which must contain an equal number of LE and non-LE images, while our implementation can tolerate class imbalance through the application of a focal loss. Furthermore, the usage of region-based CNNs is particularly suited to application on image sequences, and applying our model to time-resolved LE image sequences is an on-going project. }

All plots and associated model outputs can be find at the project GitHub repository AILE\footnote{\url{https://github.com/xiaolng/AILE}}. 

\section{Conclusion and Future Work}
\label{sec:conclusion}
In this paper, we described a novel LE detection model based on an AI object detection framework \red{applied to a} dataset of template-subtracted images from the ATLAS survey. As described in  \autoref{sec:data}, we labeled three classes: LEs \red{(}diffusive features from reflections of transients off of interstellar dust\red{)}, stars \red{(}point-like sources\red{)}, and ``other'' \red{(}a catch-all category used mostly for artifacts\red{)}.  We designed the detection model based on the architecture of YOLOv3, as described in \autoref{sec:method}. Unlike traditional object detection tasks, LEs are diverse, extended features, their shapes vary over sky positions as well as time, their edges are not sharp and blend into the background, and their structure is complex.  Thus, human labels can be quite subjective. Therefore, we modified the loss function used to train the model to incorporate both single LEs and LE-groups. We found that an \emph{F1} $\sim0.7$ can be achieved at precision 70\% even with an extremely limited training set, establishing \red{region-based} CNNs as viable detection model architectures for \red{localization and classification of LEs within an astronomical image}.

\red{This work demonstrates that modern region-based object detection architectures -- i.e., architectures that identify and then classify regions of interest as opposed to sliding window, classification-only architectures that classify every sub-region of an image -- produce detection accuracies for LEs that allow for the searching of extremely large upcoming data sets for these objects, a task which may very well be unfeasible with sliding-window classification models.  And while it is true that these region-based models are rapidly evolving (YOLO itself is now at version 7, \citealt{wang2022yolov7}, and other more recent models include \citealt{chen2019mmdetection, wu2019detectron2} for example), the core conceptual foundations are common to all such models.  That is, the models presented in this paper have demonstrated that region-based models are applicable to astronomical data (low SNR, very high spatial resolution, \emph{etc.}) for the detection of diffuse, morphologically complex features at the limit of the image SNR, suggesting that as these models are improved and new models (\eg, transformer-based architectures) are developed within the computer vision community, they will likely also be applicable to this problem when appropriate steps are taken to perform specific image preparation and definition of appropriate model architecture elements, including specialized loss functions.}

While the aim of this paper was to provide a proof-of-concept, representing the first step toward the goal of building an AI-based LE detection platform in the Rubin LSST era, we have in fact made significant progress despite limitations in the presently available data.  \red{We achieved relatively high model accuracies (\emph{F1} $\sim0.7$) with a very limited training dataset size. We demonstrated that the number of false positives scales sublinearly with the number of images \autoref{fig:false_positive} suggesting that our method can be succesfully trained on and applied to large datasets without being overwhelmed by false positives or losing detection power.  Furthermore, our model is also robust to significant class imbalance between labeled classes ($\sim$600 LEs, $\sim$1300 stars)}.  

\red{Given that the ATLAS survey has relatively few LEs with uniquely characteristic image properties, this raises the question as to the transferability of the models presented in this work.  We note however, that this is in fact a strength of our study since, while it is certainly true that any new data set will require new training and testing sets to be created and used to train the model, we have demonstrated that achieving reasonable testing accuracy is possible even with very few training examples with significant image noise and confounding non-LE artifacts and objects.  We expect then that for upcoming data sets with higher SNR and more potential LE and non-LE objects on which to train, models can only be refined further.  That is, since the models presented here are able to achieve reasonable accuracy, this provides  confidence that at least comparable (but likely much higher) accuracy on larger and cleaner datasets will be possible.}

\red{For example, the ambitious scientific goals that LSST has set to pursue \citep{ivezic2019lsst} impose unprecedented requirements on image quality, including PSF characterization, and control over systematics in the images (see for example \citealt{https://doi.org/10.48550/arxiv.1809.01669})}. Thus, the noise properties of the Rubin LSST images are expected to be far superior than current synoptic surveys, including ATLAS. By developing a model able to detect bright LEs in ATLAS images we have implicitly demonstrated that we will be able to push the detection limit to far fainter light echoes in the LSST era.  \red{Particularly, the 5-sigma limit of the survey is expected to be reached at a magnitude of $g\sim{24.5}$ and $r\sim{24.0}$,\footnote{\url{https://pstn-054.lsst.io}} enabling the detection of far fainter LEs than the ATLAS survey; the plate scale of 0.2'' will lead to seeing limited images at a site where sub-arcsecond seeing is common, enabling an accurate characterization of the shape of the PSF, leading to an LSST DIA pipeline expected to produce a minimum number of image artifacts, including in high density, high dust regions of the Galactic plane.\footnote{\url{https://pstn-039.lsst.io/}} Furthermore, the LSST survey will collect a median of 815 images for every position in the sky which will enable the creation of increasingly deeper templates over the lifetime of the survey, enabling progressive improvements in the detection of LEs in LSST.}

In future work, we will expand our training/testing datasets with both additional observational data as well as \red{additional augmentations and} simulations to increase the accuracy of the model and its performance against a variety of morphologies and to push into ever lower SNR regimes.

Two important LE features remain to be explored: their color and their time evolution.  The ATLAS data included five observing epochs, while each Rubin LSST field will be observed at least 800 times over the 10 years of the survey (with more epochs available in selected fields). The presence  of a LE in multiple images collected at different times confirms the veracity of the LE and helps removing FPs in visual inspection (for example those caused \red{by} light reflection inside of the telescope optical tube).  Images taken from different epochs also provide an opportunity to study the time evolution of LEs. Leveraging the multi-epoch nature of the ATLAS dataset will be explored in future work. Furthermore, while ATLAS provides images in a single optical band, Rubin will observe in six filters. Exploiting the color properties of LEs in their detection and classification is also part of our plan for future work.


\begin{acknowledgments}
This paper was supported by the National Science Foundation Grant No.2108841:
Detecting and studying light echoes in the era of Rubin and Artificial Intelligence NSF; and No.1814993: An Astronomical Time Machine: Light Echoes from Historic Supernovae. We acknowledge the support of the Vera C. Rubin Legacy Survey of Space Time Science Collaborations\footnote{\url{https://www.lsstcorporation.org/science-collaborations}} and particularly of the Transient and Variable Star Science Collaboration\footnote{\url{https://lsst-tvssc.github.io/}} (TVS SC) that provided opportunities for collaboration and exchange of ideas and knowledge.  This work has made use of data from the Asteroid Terrestrial impact Last Alert System (ATLAS) project. The ATLAS project is
primarily funded to search for near earth asteroids through NASA grants NN12AR55G, 80NSSC18K0284, and 80NSSC18K1575;
by products of the Near-Earth Object (NEO) search include images and catalogs from the survey area.

We used the following \red{software} packages: \texttt{astropy} \citep{astropy:2013, astropy:2018}, \texttt{numpy} \citep{harris2020array}, \texttt{pandas} \citep{mckinney-proc-scipy-2010, reback2020pandas}, \texttt{matplotlib} \citep{Hunter:2007}, \texttt{opencv} \citep{opencv_library}, \texttt{TensorFlow} \citep{tensorflow2015-whitepaper}, and \texttt{labelme} \citep{labelme2016}.

Our dataset is augmented using modules in GitHub repository \footnote{ \url{https://github.com/Paperspace/DataAugmentationForObjectDetection} }. The implementation of YOLOv3 archetecture is based on the GitHub repository \footnote{\url{https://github.com/pythonlessons/TensorFlow-2.x-YOLOv3}}

\end{acknowledgments}

\bibliography{refs}
\end{document}